# Models Coupling Urban Growth and Transportation Network Growth : An Algorithmic Systematic Review Approach

Juste Raimbault[1]

## Abstract

A broad bibliographical study suggests a scarcity of quantitative models of simulation integrating both network and urban growth. This absence may be due to diverging interests of concerned disciplines, resulting in a lack of communication. We propose to proceed to an algorithmic systematic review to give quantitative elements of answer to this question. A formal iterative algorithm to retrieve corpuses of references from initial keywords, based on text-mining, is developed and implemented. We study its convergence properties and do a sensitivity analysis. We then apply it on queries representative of the specific question, for which results tend to confirm the assumption of disciplines compartmentalisation.

## Keywords

Territory-Network Interactions ; Modeling ; Systematic Review ; Bibliometrics

## Introduction

Transportation networks and urban land-use are known to be strongly coupled components of urban systems at different scales (Bretagnolle, 2009). One common approach is to consider them as co-evolving, avoiding misleading interpretations such as the *myth of structural effect of transportation infrastructures* (Offner, 1993). A question rapidly arising is the existence of models endogeneizing this co-evolution, i.e. taking into account simultaneous urban and network growth. We try to answer it using an algorithmic systematic review. The rest of the paper is organised as follows : after a brief state of the art of existing literature, we present the approach and formalise the algorithm, which results are then presented and discussed.

[1] UMR CNRS 8504 Géographie-cités and UMR-T IFSTTAR 9403 LVMT

## Modeling Interactions between Urban Growth and Network Growth : An Overview

**Land-Use Transportation Interaction Models.** A wide class of models that have been developed essentially for planning purposes, which are the so-called *Land-use Transportation Interaction Models*, is a first type answering our research question. See recent reviews (Chang, 2006, Iacono et al., 2008), (Wegener and Furst, 2004) to get an idea of the heterogeneity of included approaches, that exist for more than 30 years. Recent models with diverse refinements are still developed today, such as (Delons et al., 2008) which includes housing market for Paris area. Diverse aspects of the same system can be translated into many models (see e.g. (Wegener et al., 1991)), and traffic, residential and employment dynamics, resulting land-use evolution, influenced also by a static transportation network, are generally taken into account.

**Network Growth Approaches.** On the contrary, many economic literature has done the opposite of previous models, i.e. trying to reproduce network growth given assumptions on the urban landscape, as reviewed in (Zhang and Levinson, 2007) . In (Xie and Levinson, 2009), economic empirical studies are positioned within other network growth approaches, such as work by physicists giving model of geometrical network growth (Barthélemy and Flammini, 2008). Analogy with biological networks was also done, reproducing typical robustness properties of transportation networks (Tero et al., 2010).

**Hybrid Approaches.** Fewer approaches coupling urban growth and network growth can be found in the literature. (Barthélemy and Flammini, 2009) couple density evolution with network growth in a toy model. In (Raimbault et al., 2014), a simple Cellular Automaton coupled with an evolutive network reproduces stylised facts of *human settlements* described by LE CORBUSIER. At a smaller scale, (Achibet et al., 2014) proposes a model of co-evolution between roads and buildings, following geometrical rules. These approaches stay however limited and rare.

## Bibliometric Analysis

Literature review is a crucial preliminary step for any scientific work and its quality and extent may have a dramatic impact on research quality. Systematic review techniques have been developed, from qualitative review to quantitative meta-analyses allowing to produce new results by combining existing studies (Rucker, 2012) . Ignoring some references can even be considered as a scientific mistake in the context of emerging information systems (Lissack, 2013). We aim to take advantage of such techniques to tackle our issue.

Indeed, observing the form of the bibliography obtained in previous section raises some hypothesis. It is clear that all components are present for co-evolutive models to exist but different concerns and objectives seem to stop it. As it was shown by (Commenges, 2013) for the concept of mobility, for which a "small world of actors" relatively closed invented a notion *ad hoc*, using models without accurate knowledge of a more general scientific context, we could be in an analog case for the type of models we are interested in. Restricted interactions between scientific fields working on the same objects but with different purposes, backgrounds and at different scales, could be at the origin of the relative absence of models of co-evolution.

While most of studies in bibliometrics rely on citation networks (Newman, 2013) or co-autorship networks (Sarigöl et al., 2014), we propose to use a less explored paradigm based on text-mining introduced in (Chavalarias and Cointet, 2013), that obtain a dynamic mapping of scientific disciplines based on their semantic content. For our question, it has a particular interest, as we want to understand *content structure* of researches on the subject. We propose to apply an algorithmic method described in the following. The algorithm proceeds by iterations to obtain a stabilized corpus from initial keywords, reconstructing scientific semantic landscape around a particular subject.

**Description of the Algorithm.** Let A be an alphabet, $A^*$ corresponding words and $T = \cup_{k \in \mathbb{N}} A^{*k}$ texts of finite length on it. A reference is for the algorithm a record with text fields representing title, abstract and keywords. Set of references at iteration *n* will be denoted $C_n \in T^3$. We assume the existence of a set of keywords $K_n$, initial keywords being $K_0$. An iteration goes as follows :

- A raw intermediate corpus $R_n$ is obtained through a catalog request, given previous keywords $K_{n-1}$.

- Overall corpus is actualized by $C_n = C_{n-1} \cup R_n$.

- New keywords $K_n$ are extracted from corpus through Natural Language Processing treatment, given a parameter $N_k$ fixing the number of keywords.

The algorithm stops when corpus size becomes stable or a user-defined maximal number of iterations has been reached. Figure 1 shows the global workflow.

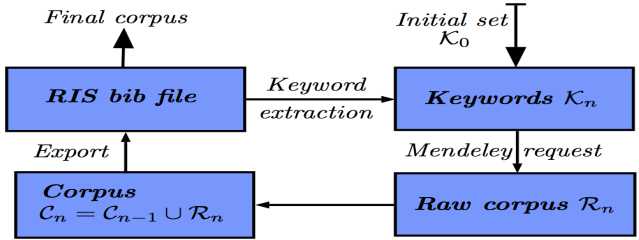

**Figure 1 : Global workflow of the algorithm.** We provide also implementation details : catalog request is done through Mendeley API ; final state of corpuses are RIS files.

## Results

***Implementation.*** Because of the heterogeneity of operations required by the algorithm (references organisation, catalog requests, text processing), it was found a reasonable choice to implement it in Java. Source code is available on the Github repository of the project[2]. Catalog request, consisting in retrieving a set of references from a set of keywords, is done using the Mendeley software API (Mendeley, 2015). Keyword extraction is done with Natural Language Processing (NLP) techniques, following the workflow given in (Chavalarias and Cointet, 2013) which is implemented by a Python script that using (NLTK, 2006).

***Convergence and Sensitivity Analysis.*** A formal proof of algorithm convergence is not possible as it will depend on the unknown structure of request results and keywords extraction. We need thus to study empirically its behavior. Good convergence properties but various sensitivities to $N_k$ were found as presented in Figure 2. We also studied the internal lexical consistence of final corpuses as a function of keywords number, defined as average dissimilarity index on cumulated co-occurrences. As expected, small keyword numbers yield more consistent corpuses, but the variability when increasing stays reasonable with a maximal increase of mean consistence around 50% (see project repository for more results on algorithm behavior exploration).

---

[2]At
github.com/JusteRaimbault/CityNetwork/tree/mas
ter/Models/Biblio/AlgoSR/AlgoSRJavaApp

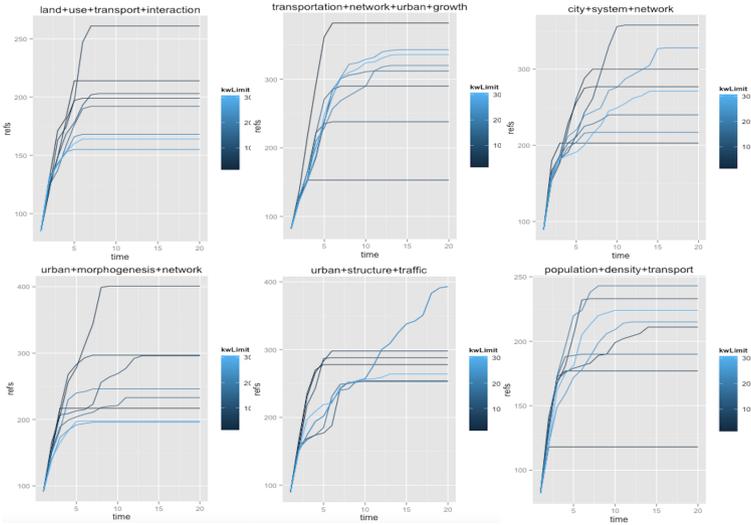

**Figure 2 : Convergence and sensitivity analysis.** Plots of number of references as a function of iteration, for various queries linked to our theme (see further), for various values of $N_k$ (from 2 to 30). We obtain a rapid convergence for most cases, around 10 iterations needed. Final number of references appears to be very sensitive to keyword number depending on queries, what seems logical since encountered landscape should strongly vary depending on terms.

**Application.** Once the algorithm is partially validated, we apply it to our question. We start from five different initial requests that were manually extracted from the various domains identified in the manual bibliography (that are "city system network", "land use transport interaction", "network urban modeling", "population density transport", "transportation network urban growth"). We take the weakest assumption on parameter $N_k = 100$ as it should less constrain reached domains. After having constructed corpuses, we study their lexical distances as an indicator to answer our initial question. Large distances would go in the direction of the assumption made in section 2, i.e. that discipline self-centering may be at the origin of the lack of interest for co-evolutive models. We show in Table 1 values of relative lexical proximity, that appear to be significantly low, confirming this assumption.

Further work is planned towards the construction of citation networks through an automatic access to Google Scholar that provides backward citations. The confrontation of corpuses modularity in the citation network with lexical proximity results are an essential aspect of a further validation of our results.

| Corpuses | 1 | 2 | 3 | 4 | 5 |
|---|---|---|---|---|---|
| 1 (W=3789) | 1 | 0 | 0.07 | 0.01 | 0.07 |
| 2 (W=5180) | 0 | 1 | 0.03 | 0 | 0.01 |
| 3 (W=3757) | 0.07 | 0.03 | 1 | 0.01 | 0.17 |
| 4 (W=3551) | 0.01 | 0 | 0.01 | 1 | 0.03 |
| 5 (W=8338) | 0.07 | 0.0125 | 0.17 | 0.03 | 1 |

**Table 1.** Symmetric matrix of lexical proximities between final corpuses, defined as the sum of overall final keywords co-occurrences between corpuses, normalized by number of final keywords (100). We obtain very low values, confirming that corpuses are significantly far.

## Conclusion

The disturbing absence of models simulating the co-evolution of transportation networks and urban land-use, confirmed through a state-of-the-art covering various domains, may be due to the absence of communication between scientific disciplines studying different aspects of these questions. We have proposed an algorithmic method to give elements of answer through text-mining-based corpus extraction. First numerical results seem to confirm the assumption. However, such a quantitative analysis should not be considered alone, but rather come as a back-up for qualitative studies that will be the object of further work, such as the one lead in (Commenges, 2013), in which questionnaires with historical actors provide relevant information on the history of modeling practices.